# A Hybrid Defense Method against Adversarial Attacks on Traffic Sign Classifiers in Autonomous Vehicles

ZADID KHAN[1], MASHRUR CHOWDHURY[1], (Senior Member, IEEE), and SAKIB MAHMUD KHAN[1], (Member, IEEE)
[1]Glenn Department of Civil Engineering, Clemson University, Clemson, SC 29634 USA

Corresponding author: Sakib M. Khan (e-mail: sakibk@clemson.edu).

This work was supported Center for Connected Multimodal Mobility ($C^2M^2$) (a US Department of Transportation Tier 1 University Transportation Center) headquartered at Clemson University, Clemson, South Carolina, US. Clemson University is acknowledged for generous allotment of computing time on the Palmetto cluster.

**ABSTRACT** Adversarial attacks can make deep neural network (DNN) models predict incorrect output labels, such as misclassified traffic signs, for autonomous vehicle (AV) perception modules. Resilience against adversarial attacks can help AVs navigate safely on the road by avoiding misclassication of signs or objects. This DNN-based study develops a resilient traffic sign classifier for AVs that uses a hybrid defense method. We use transfer learning to retrain the Inception-V3 and Resnet-152 models as traffic sign classifiers. This method also utilizes a combination of three different strategies: random filtering, ensembling, and local feature mapping. We use the random cropping and resizing technique for random filtering, plurality voting as ensembling strategy and an optical character recognition model as a local feature mapper. This DNN-based hybrid defense method has been tested for the no attack scenario and against well-known untargeted adversarial attacks (e.g., Projected Gradient Descent or PGD, Fast Gradient Sign Method or FGSM, Momentum Iterative Method or MIM attack, and Carlini and Wagner or C&W). We find that our hybrid defense method achieves 99% average traffic sign classification accuracy for the no attack scenario and 88% average traffic sign classification accuracy for all attack scenarios. Moreover, the hybrid defense method, presented in this study, improves the accuracy for traffic sign classification compared to the traditional defense methods (i.e., JPEG filtering, feature squeezing, binary filtering, and random filtering) up to 6%, 50%, and 55% for FGSM, MIM, and PGD attacks, respectively.

## I. INTRODUCTION

Autonomous vehicle (AV) software stack can be divided into perception, localization and mapping, path planning, actuation, and control based on the tasks an AV needs to perform in real-time for moving from one point to another [1]. In the perception module, the AV perceives the environment around it, including the traffic sign, using different sensors such as cameras, LiDAR, and Radar. If an AV is unable to classify traffic signs correctly due to compromised cybersecurity, it will create a significant safety issue for an AV [2]. Once an AV is connected with the external world via vehicle-to-everything (V2X) communication, the attack surface gets expanded as V2X creates opportunities for cyberattacks on the AV sensor collected data, including the traffic sign images collected through AV cameras [2], [3]. To classify traffic signs, AVs feed the image from the camera sensor to a machine learning (ML) or deep neural network (DNN) model, such as the convolutional neural network (CNN) model that is pre-trained on images of traffic signs. CNNs have made remarkable progress in object recognition, surpassing that of humans in the ImageNet Large Scale Visual Recognition Challenge in 2017 with an error rate of 2.25% [4]. However, research has shown that many security threats can compromise a DNN model, including a CNN model, which make it predict or classify objects incorrectly. Among the threats, adversarial attacks have shown to be successful in speech recognition [5], robot vision [6] and image classification [7]. In an adversarial attack scenario, an attacker creates a perturbation model to alter the input (such as a traffic sign image) and make the DNN model predict incorrect traffic sign labels. Placing small stickers or patches on traffic signs can make the DNN model misclassify the image label as well. Previous studies have already shown that adversarial attacks can deceive a traffic sign classifier in vehicles [8]–[10]. The perturbation may be minimal such that the alterations are not visible to the human eye, but these small perturbations are sufficient to deceive a deep learning image classifier. Our goal is to develop a resilient traffic sign classifier for AVs. Following [11], we define a resilient classifier as one that can maintain its classification performance even after an adversarial attack has impacted the input data.

There are primarily two types of adversarial attacks; untargeted and targeted. In the untargeted attack, the perturbation model is designed to make the DNN model misclassify without any target label. For the targeted attack, the perturbation model is designed so that the DNN model predicts a specific target label. From another viewpoint, there are three types of adversarial attacks; white-box, gray-box, and black-box attacks [12]. For the black-box attack, the attacker does not know the DNN model, its parameters, or the defense method in use. In this case, the attacker creates the perturbation model based on some generic model and relies on the transferability property. The attacker knows the DNN model and its parameters in the gray-box setting, but it has no access to the defense method. In a white-box attack, the attacker has full information about the model, its parameters and the defense method. All these attacks can be targeted or untargeted depending on the perturbation model. Considering the impact of adversarial attacks on traffic sign classifiers for AVs, it is essential to develop resilient defense methods for traffic sign classifiers under adversarial attacks.

Some popular defense methods against adversarial attacks are adversarial training, adversarial detection, input reconstruction, input denoising, classifier robustification, network verification and a combination of multiple models, such as defensive distillation and ensembling [13]. While these methods can increase the resilience of ML models in attack scenarios, their performances are not often comparable to the non-attack scenarios. In [14], authors evaluated adversarial training and defensive distillation methods to detect adversarial attacks (i.e., spatial attacks and Carlini and Wagner or C&W attacks) for recognizing colors of traffic lights through image classfication. Compared to the no-defense scenario, the authors in [14] were able to get a lower attack success rate in only one scenario (spatial attack detection with adversarial training) among all evaluation scenarios. Our study focuses on developing a resilient traffic sign classification system that performs similarly (e.g., maintain similar accuracy in traffic sign detection) regardless of attack or no-attack scenarios for different attack scenarios.

In this stufy, we develop a hybrid defense method combining three different approaches: random filtering, ensembling and local feature mapping. We focus on an untargeted white-box setting since it allows the attacker full access to the DNN model.

We first discuss the study contributions (Section II) and literature on adversarial attack detection and mitigation (Section III). Section IV and V describe the attack models and hybrid defense method, respectively. Section VI and VII describe the research method and our analysis findings, respectively. Section VIII discusses the conclusions based on the results obtained in Section VII and provides recommendations for future research.

## II. CONTRIBUTIONS OF THIS STUDY

Our DNN-based hybrid adversarial defense approach, which includes random filtering, ensembling and local feature mapping, is the first of its kind, and no previous work has attempted such a hybrid method for defense against adversarial attacks to the sign recognition of the AV perception module. Moreover, this is the first formal defense method specific to adversarial attacks on traffic sign classifiers, as previous models have been tested on generic datasets [12], [15], [16]. Previous literature has presented deterministic defense methods that are not resilient to new attack types beyond the attacks considered in the research [14]. Our defense method achieves similar accuracies in attack and non-attacks scenarios, and it is not specific to any particular type of attack. The presence of multiple DNN models makes it difficult for attacks to find an optimum adversary that can deceive all models. Moreover, random filtering introduces randomness in the classification process and localized features increase model resilience since these features are less susceptible to adversarial perturbations applied to the entire image.

## III. RELATED WORKS ON MITIGATING ADVERSARIAL ATTACKS AND KNOWLEDGE GAPS

Defense against adversarial attacks and improving deep learning model resilience is a growing research area. Many recent studies have developed defense methods by combining different strategies. Here, we describe some state-of-the-art defense methods and the knowledge gaps we will address in our study.

### A. DEFENSE METHODS FOR MITIGATING ADVERSARIAL ATTACKS

#### 1) ADVERSARIAL TRAINING

Adversarial training is a defense method where the classification model is retrained using adversarial examples [13]. It has been shown that retraining the models using adversarial samples from powerful models, such as Projected Gradient Descent or PGD, is an effective way to increase the model resilience against many types of adversarial attacks [15], [17]. Many existing studies have used adversarial training as a defense method [18], [19].

### 2) GENERATIVE ADVERSARIAL NETWORKS

Generative adversarial networks (GAN) are also popular defense methods [17]. For example, Yuan and He developed an adversarial dual network learning method supported by random image transform as a defense method [16]. At first, the input image is transformed to override the noise of an adversarial attack. After the image transformation, a GAN model is used, which contains two different sub-models; a generative cleaning model and a detector model. The generative cleaning model reconstructs the original image by removing the initial noise from the image transformation. The detector model determines whether the input image is under attack. The generative cleaning model and detector model are trained together using adversarial learning. This adversarial dual network learning method performs well compared to baseline adversarial learning methods for different white, black and grey box attacks. However, the accuracy under attack is still significantly lower than accuracy with no attack. Also, this adversarial dual network learning method is not efficient because it has high computational requirements and does not protect against all types of perturbations. New types of attack models can be introduced that can generate adversarial samples and bypass any existing defense method [15].

### 3) DENOISER MODEL

Denoising input images before feeding them to the classification model is an effective defense method. In this method, a denoiser model, such as an autoencoder, can be used to remove the noise in the image, which would increase the classification accuracy [15]. Several studies have used denoising models for adversarial resilience [12], [20]. For example, a defense method was developed by Bakhti et al. with the deep denoising sparse autoencoder (DDSA) [12]. DDSA removes the noise from the image before it enters the classification model. The denoising autoencoder is used for dimensionality reduction and reducing the effect of noise; the sparsity constraint is used for sparse activation in the fully connected layers of the autoencoder, thus enabling the neurons to be intermittently inactive and extracting meaningful and relevant features. The model is trained using clean data and adversarial examples using PGD attacks. Although it has been proven that PGD samples generalize well for all types of attacks, they cannot cover all types of attacks, so an autoencoder trained on PGD samples is not a universal and generic defense method [15]. Another method for denoising is feature denoising. Instead of denoising at the pixel level, a model can be developed that removes the noise from the extracted feature space introduced by adversarial attacks [17]. In both cases, the denoiser network needs adversarial samples for training, so it suffers from the same issue as adversarial training [13].

### 4) IMAGE TRANSFORMATION AND FILTERING

Image transformation and filtering are effective methods against adversarial attacks [13]. It is a generic process where some transformation is applied to the input to reduce the effect of perturbation [17]. Several studies have used image filtering techniques to increase model accuracy [21], [22]. Aprilpyone et al. developed an adversarial defense method using one-bit double quantization process (i.e., mapping process to reduce the input values) and Resnet models [21]. First, 1-bit dithering was applied on training images.. In the testing phase, 1-bit dithering was applied to the input image, followed by a linear quantization stage for removing the noise. Three different white-box attacks were created on two different datasets and showed improvement over traditional defense methods. However, the limitation of [21] is that their defense method cannot handle attacks on the image before 1-bit dithering. This is not practical in real-world scenarios, where it is not known whether an image is clean or noisy [21]. The advantage of this method is that it is not specific to certain types of attacks. It is a generic approach that can be applied to any scenario. However, this method may reduce the accuracy of the no-attack scenario [17].

### 5) ENSEMBLING

Another popular technique for making the DNN model resilient against adversarial attacks is ensembling [13]. It is difficult to create an attack model that can deceive all models within the ensembled attack detection model. That is why an ensemble approach with an aggregation strategy to combine the model outputs is an effective defense method [15]. Several studies have explored ensembling techniques [23], [24]. For example, Mun and Kang developed an ensemble method for achieving adversarial resilience. They encoded the output of the CNN model using a random binary method and created an ensemble of CNN models with different encoding schemes [23]. The ensemble of models is trained together using a unified global objective function. They have used three different types of white-box attacks on three benchmark classification datasets and showed that using an ensemble of models with different encoding schemes improves the classification accuracy compared to using the same encoding scheme. However, the authors in [23] did not compare their method with other state-of-the-art adversarial defense methods. The models that are part of the ensemble need to be independently effective classifiers; otherwise, the model's overall accuracy decreases.

### B. KNOWLEDGE GAPS IN EXISTING LITERATURE

Overall, from the literature review, we have found several gaps for resiliency against adversarial attacks which we address in this study. At first, existing defense methods have been tested for benchmark datasets such as MNIST (contains images of hand-written digits), Fashion-MNIST (contains images of fashion accessories and clothes), and CIFAR-10 (contains images of airplane, automobile, bird, cat, deer, dog, frog, horse, ship, and trucks) [12], [15], [16]. The effect of the adversarial attack on traffic sign classifiers and potential defense methods is still greatly an unexplored area. Also, the investigation about a generalized defense method that can be effective against all adversarial attacks is missing. No defense method used local feature mapping to increase DNN model resilience. Finally, the combination of random filtering, ensembling, and local feature mapping has not been explored as a defense strategy for adversarial attacks against DNN-based object recognition in previous studies.

## IV. ATTACK MODELS

Here we discuss different types of white-box attacks which are studied in the literature [12], [25]. These attack models are used in our experiments.

### A. FAST GRADIENT SIGN METHOD (FGSM) ATTACK

Goodfellow et al. first proposed FGSM (an untargeted attack) to create adversarial samples [26]. FGSM creates the attack along the direction of the gradient of the adversarial loss. The FGSM-generated adversarial sample is generated using Eq. (1).

$$x' = x + \varepsilon.sign[\nabla_x J(\theta, x, y)] \quad (1)$$

Where $y$ is output, $x$ is input before the attack, $\theta$ is the set of parameters of the model, $x'$ is the input after the attack, $\varepsilon$ is the magnitude of the perturbation, $\nabla_x$ is the differential operator with respect to $x$, and $J$ is the loss function (cost). Figure 1 shows the effect of the FGSM attack on a sample stop sign image. The individual pixel values have been changed under this atatck so that if the captured image still looks like a stop sign, a classifier misclassifies it as a merge sign.

### B. MOMENTUM ITERATIVE METHOD (MIM) ATTACK

Dong et al. developed a new iterative algorithm called MIM combining the momentum memory and the basic iterative method (BIM) [27]. Here, momentum is a part of the gradient descent method. In gradient descent, the direction of the steepest slope is calculated for a function, and a momentum term ensures that the algorithm does not change direction instantly at some point of the function; rather, it keeps going in the same direction for some time before changing direction. Specifically, MIM iteratively updates the adversarial sample based on Eq. (2).

$$x'_{t+1} = x'_t + \alpha.sign(g_{t+1}) \quad (2)$$

Here, Eq. (3) is used to update the gradient $g$.

$$g_{t+1} = \xi.g_t + \frac{\nabla_x J(\theta, x_t, y)}{\|\nabla_x J(\theta, x_t, y)\|} \quad (3)$$

Here, $\alpha$ is the magnitude of the perturbation, and $\xi$ is a decay factor. Since it is an iterative process, the perturbation for the next time step $t+1$ depends on the perturbed input $x_t'$ and gradient $g_t$ from the current timestep $t$. The combination of MIM and ensemble attack strategy won both targeted and non-targeted adversarial attack competitions (black-box) at the Neural Information Processing Systems conference in 2017 [27]. Figure 1 shows the effect of the MIM attack on a sample stop sign image. The individual pixel values have been modified such that it still looks like a stop sign, but a classifier misclassifies it as a merge sign.

### C. PROJECTED GRADIENT DESCENT (PGD) ATTACK

The PGD can be considered as a generalized version of BIM without the constraint on the perturbation value ($\varepsilon$). PGD constrains the adversarial perturbations by projecting the learned adversarial samples into the $\varepsilon$-$L_{inf}$ neighbor of the benign samples [18]. The adversarial perturbation size in PGD is lower than $\varepsilon$. Eq. (4) is used for the update procedure.

$$x'_{t+1} = Proj\{x'_t + \alpha.sign[\nabla_x J(\theta, x_t, y)]\} \quad (4)$$

Where *Proj* projects the updated adversarial sample into the $\varepsilon$-$L_{inf}$ neighbors which are still within a valid range. Figure 1 shows the effect of the PGD attack on a sample stop sign image. The individual pixel values have been modified in a way that it looks like a stop sign, but a classifier misclassifies it as a merge sign.

### D. CARLINI AND WAGNER (C&W) ATTACK

Carlini and Wagner proposed C&W attacks, which are optimization-based adversarial attacks. C&W attacks create $L_0$, $L_2$, and $L_{inf}$ norm measured adversarial samples, which are denoted as $CW_0$, $CW_2$, and $CW_{inf}$ [28]. In the C&W attack, the optimization objective is formulated as follows:

$$\min_\delta D(x, x+\delta) + c.f(x+\delta) \quad (5)$$

Where $x + \delta \in [0,1]$; $D$ is the $L_0$, $L_2$, or $L_{inf}$ distance metric; and $\delta$ is the adversarial perturbation. $f(x+\delta)$ means a adversarial loss that satisfies $f(x+\delta) <0$ when the DNN predicts an attack target. C&W attack model introduces variable $\kappa$ to substitute $\delta$ using Eq. (6) so that $(x+\delta)$ yields a valid image.

$$\delta = \frac{1}{2}[\tanh(\kappa) + 1] - x \quad (6)$$

which ensures that $(x+\delta)$ is always in the range [0,1] in the optimization process. C&W attacks are very powerful for generating adversarial examples. They successfully impacted DNN's performance while MNIST, CIFAR-10, and ImageNet datasets were used [28]. They compromised defensive distilled models, such as Limited-memory Broyden–Fletcher–Goldfarb–Shanno and DeepFool, and prevented these models from accurately identifying the adversarial samples [28]. Figure 1 shows the effect of the C&W attack on a sample stop sign image. The individual pixel values have been changed in a way that it still looks like a stop sign, but a classifier misclassifies it as a merge sign.

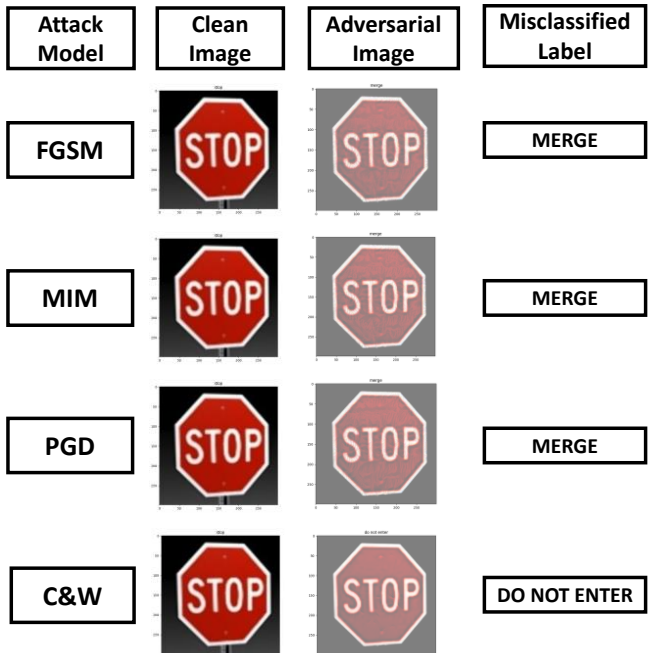

FIGURE 1. Effect of attack models and image transformations on sample traffic sign image (STOP sign).

## V. HYBRID DEFENSE METHOD

This section describes the hybrid defense method, considered in this study, for improving the resilience of traffic sign classification regardless of the type of adversarial attack. The hybrid defense method is developed based on two DNN models: Inception-V3 and Resnet-152. The original DNN models are retrained with transfer learning. Later the retrained DNN models are combined with three different defense strategies to improve the model resilience against adversarial attacks. The three strategies are: random filtering, ensembling, and local feature mapping. We refer to the combined model (retrained DNN incorporating three additional strategies) as the hybrid defense method. The steps of the defense method are shown in Figure 2 and discussed in the following.

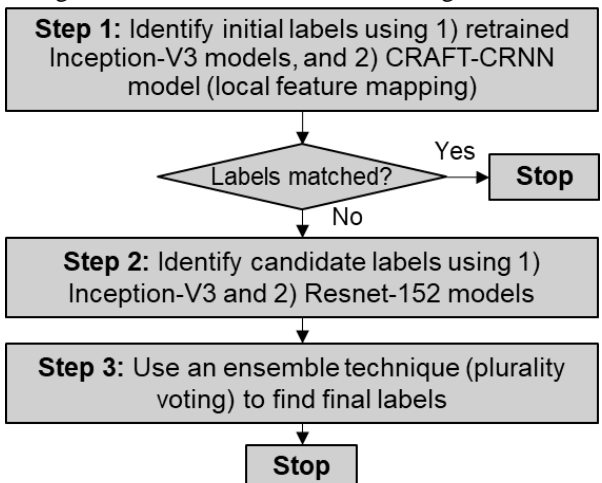

FIGURE 2. Steps of the hybrid defense method.

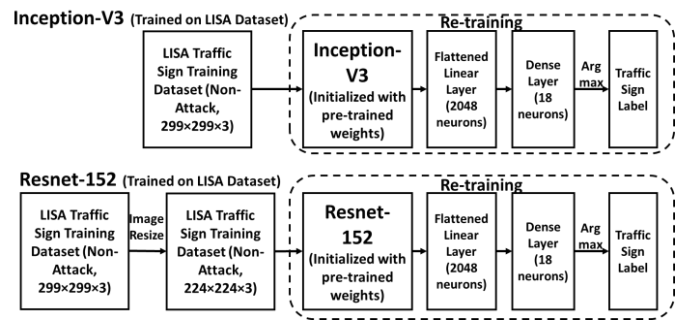

FIGURE 3: Transfer learning applied to Inception-V3 and Resnet-152 model.

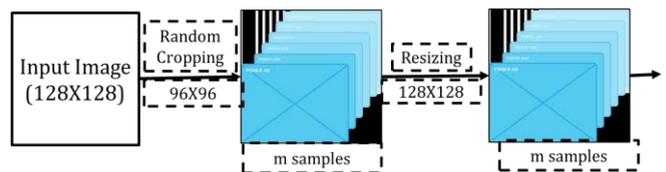

FIGURE 4. Random cropping and resizing applied to a sample input image.

### A. STEP 1: IDENTIFY INITIAL LABELS USING INCEPTION-V3 MODEL AND LOCAL FEATURE MAPPING

The details of the transfer learning-based Inception-V3 classifier and local feature mapping used in this study are given below.

#### 1) LABEL WITH RETRAINED INCEPTION-V3 MODEL

The base model for traffic sign classification is the Inception-V3 model. The Inception-V3 model is the third version of Google's CNN model primarily created for the ImageNet recognition challenge. It is an improvement to the previous inception models focusing on fewer computational requirements. Several techniques have been used to optimize network performance, such as factorized convolutions, regularization, dimension reduction, and parallelized computations. Figure 3 shows the transfer learning process of the Inception-V3 model. The transfer learning process is used to retrain the Inception-V3 model using traffic sign images from the LISA dataset.

While retraining the model, we used random filtering, which is a type of image transformation where some parameters of the transformation technique are generated randomly so that the adversary is unable to predict the type of transformation that the image will go through, thus weakening the effect of adversarial effect on the image classification model. In this study, we have used random cropping and resizing as the random image transformation method. In this technique, the frame size of the cropping remains fixed, but the area or position of the cropping is determined randomly. During the training of the image classification models, data augmentation is used to generate different variations of the dataset. As we use random cropping and resizing, an Inception-V3 model trained using image augmentation is able to classify images with cropped versions of corresponding images. This means that the model is trained to predict using partial information from images. This capability of the Inception-V3 model is used to improve the model resilience in this study. Figure 4 shows an example of random cropping and resizing on a 128×128 input image. The image is randomly cropped $m$ times, creating $m$ different samples from the same image, all with the same dimension of 96×96. After that, the images are resized to the dimension of 128×128. As a result, we get $m$ randomly generated samples of the same dimension (128×128) from a single input image.

Figure 3 shows the retraining process using traffic sign images. After applying transfer learning, we get an Inception-V3 model which is trained on traffic sing images, so it can be used as a standalone classifier. From Table 1, we observe the architecture of the original Inception-V3 model. The model takes RGB images of 299×299×3 dimensions as input. After that, there is a series of convolution, pooling, and inception layers. We stop at the second from the last layer, which is a flattened linear layer consisting of 2048 neurons. We add a fully connected (FC) layer consisting of 18 neurons for the 18 class labels of traffic signs. We use the argmax function to convert the logits to a single class label. This is the baseline model for traffic sign classification in this study, and we compare the performance of all other models, which include the traditional defense methods and our hybrid defense mehod, to this model. Also, as it is considered the base model for the hybrid method, the white-box attacks we generate are based on this particular model.

**TABLE 1.** Inception-V3 model architecture.

| Layer Type | Patch Size/Stride | Input Dimension |
|---|---|---|
| Conv | 3×3/2 | 299×299×3 |
| Conv | 3×3/1 | 149×149×32 |
| Conv (Padded) | 3×3/1 | 147×147×32 |
| Pool | 3×3/2 | 147×147×64 |
| Conv | 3×3/1 | 73×73×64 |
| Conv | 3×3/2 | 71×71×80 |
| Conv | 3×3/1 | 35×35×192 |
| 3XInception | N/A | 35×35×288 |
| 5XInception | N/A | 17×17×768 |
| 2XInception | N/A | 8×8×1280 |
| Pool | 8×8 | 8×8×2048 |
| Linear | Logits | 1×1×2048 |
| SoftMax | Classifier | 1×1×1000 |

#### 2) LABEL WITH LOCAL FEATURE MAPPING

We examined the traffic sign dataset and found that an essential local feature of the traffic sign images is words/text. Some traffic signs contain specific text in a localized region of the image, and this text is usually unique to the traffic sign (i.e., "STOP" for stop sign). The localized text is a resilient feature since the perturbation is applied to the whole image and is not concentrated on adding noise to the text. Adversarial attacks create perturbations in images that are usually not visible to the human eye. Text detection and recognition models are more robust to variations in text patterns; hence, they effectively classify images of traffic signs with text. However, this approach is not effective for traffic signs without text. Therefore, it can be used in conjunction with other approaches, but not as a standalone defense approach. In this study, we have used the Keras optical character recognition (OCR) pipeline for extracting texts from images [29]. In this pipeline, there are two models. The CRAFT (Character Region Awareness for Text) detection model detects the regions containing the text [30]. The pipeline uses a convolutional recurrent neural network (CRNN) model for recognizing the text in the identified regions. The CRAFT model is trained on several datasets, including SynthText, IC13, IC15, and IC17. The CRNN model is trained using the Synth90K dataset. We do not re-train the model with the traffic sign dataset since the pre-trained model achieves an accuracy of 99% in detecting text in non-attack traffic sign images containing text.

The output of the Inception-V3 model is a traffic sign label. The same image is also input to the pre-trained CRAFT and CRNN models for text detection and recognition. The output is a set of words and regions from the image. A word template matching function matches the output with a set of templates and generates another traffic sign label based on template matching. A template match indicates the successful identification of relevant texts in the image. The matching of traffic sign labels means that it is the correct label and there is no attack detected. In case of a mismatch, the label generated from the CRAFT and CRNN models is trusted as the correct label. This situation is identified as the attack scenario. Failure to identify a template match means that the CRAFT and CRNN models could not detect the text, or we have encountered a traffic sign with no text, and we go to the next step.

### B. STEP 2: IDENTIFY CANDIDATE LABELS USING DNN MODELS

If labels generated from the Inception-V3 model and other models used in local feature mapping (CRAFT and CRNN models) in the previous step do not match, we conduct this another label identification using two DNN models: Resnet-152 and Inception-V3. The Resnet-152 is another traffic sign classification model besides the Inception-V3 model. In this model, residual nets having a depth of up to 152 layers are used. Table 2 shows the architecture of the Resnet-152 model.

**TABLE 2:** RESNET-152 model architecture.

| Layer Name | Output Size | Operations |
|---|---|---|
| Conv1 | 112×112 | 7×7, 64, Stride 2 |
| Conv2_x | 56×56 | [1×1×64] ×3 |
|  |  | [3×3×64] ×3 |
|  |  | [1×1×256] ×3 |
| Conv3_x | 28×28 | [1×1×128] ×8 |
|  |  | [3×3×128] ×8 |
|  |  | [1×1×512] ×8 |
| Conv4_x | 14×14 | [1×1×256] ×36 |
|  |  | [3×3×256] ×36 |
|  |  | [1×1×1024] ×36 |
| Conv5_x | 7×7 | [1×1×512] ×3 |
|  |  | [3×3×512] ×3 |
|  |  | [1×1×2048] ×3 |
| Average Pooling | 1×1 | 1000-d fc, SoftMax |

The model is divided into five stages of convolution. Each stage contains multiple layers, as shown in Table 2. For example, the second convolution stage contains nine layers. The total number of layers from the five convolution stages and the average pooling stage is 152; hence the name is Resnet-152. The input is an RGB image with dimensions 224×224×3. The output of the 5th convolution stage is a flattened linear layer consisting of 2048 neurons. Similar to the Inception-V3, we use an FC layer with 18 neurons and the argmax function to convert the logits to a single class label. We also use transfer learning to retrain the Resnet-152 model on the traffic sign dataset created in this study. Figure 3 shows the transfer learning process for the Resnet-152. The model is retrained on the traffic sign images that are used in our research. After the retraining process is complete, we obtain a modified Resnet-152 model that can be used as a standalone classifier. Random filtering is used for both Resnet-152 and Inception-V3 models.

### C. STEP 3: IDENTIFY FINAL LABEL WITH ENSEMBLING

In this study, we have developed two separate DNN-based traffic sign classification models and used an ensemble technique to combine their outputs. Random cropping and resizing are used to generate several random samples from one input image. The two classifiers (i.e., Inception-V3 and Resnet-152) are used to predict several output labels. The plurality voting technique is used to extract one output label from a list of output labels. In the ensemble technique, multiple output labels from multiple classifiers are combined into a single output.

### VI. RESEARCH METHOD

This section discusses the research method, specifically the dataset, hybrid model for the testing scenario, and traditional models for defense against adversarial attacks for traffic sign classification.

#### A. DATASET

In this study, we consider a dataset containing different types of US traffic signs. From the literature, we have found that the extended LISA traffic sign dataset contains the most comprehensive collection of US traffic sign images [31]. The traffic sign images are extracted from video frames in the LISA dataset. The video was collected from the dashboard camera of many different vehicles. The vehicles were driven around San Diego in California. The resolution of the original video frames varied between 640×480 pixels and 1024×522 pixels. The annotations of traffic signs within the frames varied between 6×6 pixels and 167×168 pixels. The images were a mixture of color and grayscale. There is a total of 7855 annotations on 6610 images. There are 47 types of traffic signs in the original dataset.

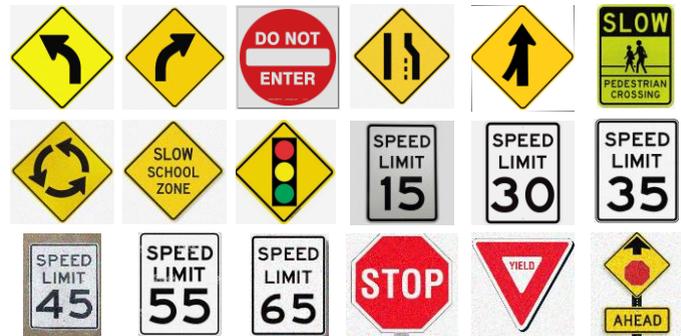

FIGURE 5. Samples of modified traffic sign dataset for this study.

After examining the dataset, we find that not all traffic sign images contain sufficient data for training and testing a classifier. Moreover, some images have the traffic sign in a small region of the whole image. Therefore, we create a modified traffic sign dataset containing only 18 types of traffic signs instead of 47 present in the original LISA traffic sign dataset for our study. We apply cropping to remove the surrounding noise and focus on the traffic sign. Figure 5 contains a sample image for each type of traffic sign from the dataset prepared in this study.

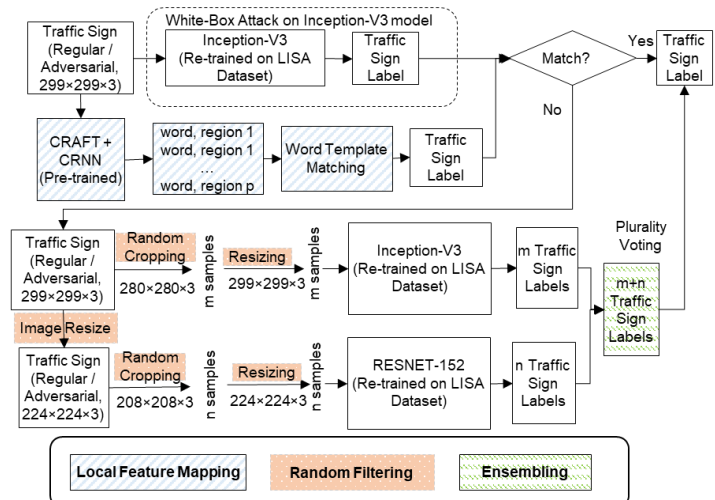

Figure 6. Hybrid defense method testing for regular and adversarial input conditions.

#### B. TESTING FOR REGULAR AND ADVERSARIAL INPUT WITH HYBRID DEFENSE METHOD

Figure 6 shows the process of the hybrid defense method for regular and adversarial input images. The adversarial attack is applied to the Inception-V3 model as it is considered the baseline classification model in this study.

The 299×299×3 image is converted to a 224×224×3 image for the Resnet-152 model. Random cropping and resizing are applied to both images simultaneously. For the Inception-V3 model, the 299×299×3 image is cropped m times randomly. The cropping dimension is kept constant, which is 280×280×3, but the cropping position is random. The 280×280×3 images are resized again to have the dimension as 299×299×3. The m 299×299×3 images are input to the Inception-V3 model, which produces m traffic sign labels. The same process is followed for the Resnet-152 model. The number of samples is changed to $n$, and the cropping dimension is changed to 208×208×3. The traffic sign labels are combined to form a traffic sign label vector of length ($m+n$). We apply plurality voting to identify the label with the highest number of occurrences. In the case of a tie, a label is chosen randomly. This label is the final output of the hybrid defense method. In this study, we have used $m=8$ and $n=10$ as these values give the highest accuracy for traffic sign classification.

### C. TRADITIONAL DEFENSE METHODS

The traditional defense methods [32]–[34] we have considered in this study are given below.

#### 1) JPEG FILTERING

JPEG compression is a popular defense method against adversarial attacks. The human visual system is less sensitive to high frequency components, and adversarial attacks aim to introduce perturbations which are imperceptible to the human eye [33]. JPEG compression suppresses the high frequency components of an image, such as sharp changes in intensity and color hue, using discrete cosine transform [34]. As a result, JPEG compression has the potential to counteract the perturbations on the original images. Compression rate is used to control the output image quality, which also has an impact on the effectiveness of JPEG compression as a defense mechanism. JPEG compression makes the pixels more coarse-grained compared to the original image. In this study, we have used a value of 50% as the quality of the JPEG filtered image, since other studies show that this value yields better effectiveness as a defense mechanism [33]. Figure 2 shows the effect of JPEG compression on a sample adversarial stop sign image.

#### 2) FEATURE SQUEEZING

Feature squeezing refers to reducing the color bit depth of every pixel in an image. It is also known as bit squeezing since reducing the color depth means reducing the number of bits required to represent a pixel's color value. The feature space reduction is helpful for defending against adversarial attacks since it converts various feature vectors in the original space into similar samples. In this study, we have used a bit depth of 4 [32] as this value produces the highest sign classification accuracy under adversarial attack for our dataset.

#### 3) BINARY FILTERING

This is the most straightforward defense method. It directly converts the input image into a binary black and white image. This is a special case of feature squeezing, where the bit depth is 1 [32]. By reducing feature space, binary filtering helps to defend against adversarial attacks since it converts various feature vectors into similar samples.

#### 4) RANDOM FILTERING

Although our hybrid method uses random filtering as a part of the defense, we want to check how random filtering performs independently as the defense method for adversarial attacks. Since we perform random filtering for the original Inception-V3 model, we crop 299×299×3 images to 280×280×3 images and resize the images again to 299×299×3 images. We create eight samples from each image and predict the class label eight times. After that, plurality voting is used to get the final class label.

## VII. ANALYSIS AND FINDINGS

This section presents the analysis and findings from the hybrid defense method for different white-box attacks. We also compare the results with other traditional defense methods. We explore class-wise performance metrics and overall metrics for the traditional methods and the hybrid method. We have used Pytorch package to implement the models and advertorch package to create the attacks on the images [35]. All models are trained and tested in the Clemson University Palmetto Cluster nodes containing Nvidia Tesla p100 GPUs [36], however in real-world the inference (i.e., classifying the images under adversarical attack) based on the hybrid defense method will happen using the in-vehicle computation units of the AVs. One of the recent computation solutions for AVs is Nvidia DRIVE AGX Pegasus, which can achieve 320 TOPS (trillion operations per second) [37]. The P100 GPU used in our experiments has a capability of 18.7 Tera Flops (floating point operations) per second, which is significantly less than the capability of the DRIVE AGX platform [38], Therefore, the hybrid defense method can be used by AVs for real-world operations.

### A. HYBRID DEFENSE METHOD PERFORMANCE EVALUATION

We compare our hybrid method with other traditional methods and the baseline model for different types of attacks. We use precision, accuracy, recall, and F1-value to compare the models. Accuracy is computed globally for all 1451 samples in the test set. For precision, recall, and F-1 value, we compute the class-wise values and use the weighted average method to get a global value. As the number of samples is different for different traffic signs, we have used a weighted average to account for the imbalance in the dataset. The results from our hybrid defense method are highlighted in gray in the tables.

**TABLE 3:** Comparison of defense methods for no attack.

| Defense Method | Precision | Recall | F1-Score | Accuracy |
|---|---|---|---|---|
| Inception-V3 | 0.98 | 0.98 | 0.98 | 0.98 |
| JPEG Filter | 0.91 | 0.88 | 0.89 | 0.88 |
| Feature Squeeze | 0.98 | 0.98 | 0.98 | 0.98 |
| Binary Filter | 0.98 | 0.98 | 0.98 | 0.98 |
| Random Filter | 0.99 | 0.99 | 0.99 | 0.99 |
| Hybrid | 0.99 | 0.99 | 0.99 | 0.99 |

#### 1) NO ATTACK

We have evaluated all adversarial defense methods in the no-attack scenario. A common pitfall of existing defense methods against adversarial attacks is that it decreases the accuracy of the classification model when the adversarial defense is added to the model in the non-attack scenario. From Table 3, it can be observed that almost all the defense methods are performing better or equal to the original Inception-V3 model, except for the JPEG filtering method, which reduces the accuracy to 88% from the Inception V3's 98% accuracy. Hybrid and random filtering methods are able to achieve accuracy, precision, recall, and F1-score of 99%.

**TABLE 4: Comparison of defense methods for FGSM attack.**

| Defense Method | Precision | Recall | F1-Score | Accuracy |
|---|---|---|---|---|
| Inception-V3 | 0.81 | 0.75 | 0.76 | 0.75 |
| JPEG Filter | 0.82 | 0.76 | 0.77 | 0.76 |
| Feature Squeeze | 0.81 | 0.75 | 0.76 | 0.75 |
| Binary Filter | 0.86 | 0.66 | 0.71 | 0.66 |
| Random Filter | 0.88 | 0.84 | 0.85 | 0.84 |
| Hybrid | 0.91 | 0.89 | 0.89 | 0.89 |

#### 2) FGSM ATTACK

The FGSM attack is implemented for ε=0.1 as suggested in the literature [39]. The results are presented in Table 4. The FGSM attack is not as effective as PGD and MIM adversarial attacks. It only reduces the accuracy of the Inception-V3 model to 75%. The hybrid method performs best with the highest precision, recall, F1-score, and accuracy of 91%, 89%, 89%, and 89%, respectively, among the other models against the FGSM attack. The random filtering improves the accuracy from 75% to 84%, and the additional increase in accuracy can be attributed to the combination of ensembling and local feature mapping. The local feature mapping is mostly responsible for classification of traffic signs with text. For non-text signs, the Resnet-152 model helps in producing the correct traffic sign label. Therefore, we can conclude that ensembling, random filtering and local feature mapping improve adversarial resilience. Other traditional methods, such as JPEG filtering and feature squeezing, are unable to improve on the base accuracy of 75%. The binary filtering degrades the performance of the Inception-V3 model. The hybrid defense method improves the accuracy by 6% over the next best traditional defense method, which is random filtering.

**TABLE 5: Comparison of defense methods for MIM attack.**

| Defense Method | Precision | Recall | F1-Score | Accuracy |
|---|---|---|---|---|
| Inception-V3 | 0.29 | 0.10 | 0.12 | 0.10 |
| JPEG Filter | 0.41 | 0.23 | 0.26 | 0.23 |
| Feature Squeeze | 0.32 | 0.12 | 0.16 | 0.12 |
| Binary Filter | 0.71 | 0.56 | 0.59 | 0.56 |
| Random Filter | 0.65 | 0.53 | 0.55 | 0.53 |
| Hybrid | 0.91 | 0.87 | 0.88 | 0.87 |

#### 3) MIM ATTACK

For the MIM attack, we have used cross-entropy as a loss function. The MIM attack is implemented for ε=0.1 [40]. The results are presented in Table 5. The MIM attack is effective in reducing the accuracy of the Inception-V3 model to only 10%. The hybrid method performs best with the highest precision, recall, F1-score, and accuracy of 91%, 87%, 88%, and 87%, respectively. Random filtering improves the accuracy from 10% to 53%, and the additional increase in accuracy can be attributed to the combination of ensembling and local feature mapping. Local feature mapping is applicable for the traffic signs with text. For non-text signs, the Resnet-152 model helps in producing the correct traffic sign label. Therefore, we can conclude that ensembling, random filtering, and local feature mapping have a noticeable impact in terms of adversarial resilience. Binary filtering performs very well in this scenario, as it has achieved 56% accuracy, which is better than the random filtering method. However, binary filtering suffers from lower accuracy for images which were not attacked, which we have addressed in the hybrid method. The hybrid method improves the accuracy by 50% over the binary filtering.

**TABLE 6: Comparison of defense methods for PGD attack.**

| Defense Method | Precision | Recall | F1-Score | Accuracy |
|---|---|---|---|---|
| Inception-V3 | 0.27 | 0.08 | 0.11 | 0.08 |
| JPEG Filter | 0.40 | 0.24 | 0.26 | 0.24 |
| Feature Squeeze | 0.31 | 0.11 | 0.15 | 0.11 |
| Binary Filter | 0.72 | 0.57 | 0.60 | 0.57 |
| Random Filter | 0.69 | 0.58 | 0.59 | 0.58 |
| Hybrid | 0.90 | 0.87 | 0.88 | 0.87 |

#### 4) PGD ATTACK

For the PGD attack, we have used the $L_{inf}$ norm and cross-entropy as loss functions. Here we examine the effect of the PGD attack for ε=0.1 [40]. The results are presented in Table 6. The PGD attack is very effective in reducing the accuracy of the Inception-V3 model to 8%. The hybrid method performs best with the highest precision, recall, F1-score, and accuracy of 90%, 87%, 88%, and 87%, respectively. Random filtering improves the accuracy from 8% to 58%, and the additional increase in accuracy can be attributed to the combination of ensembling and local feature mapping. Local feature mapping is primarily responsible for improving accuracy of the traffic signs with text. For non-text signs, the Resnet-152 model helps to generate the correct traffic sign label. Therefore, we can conclude that ensembling, random filtering, and local feature mapping significantly impact adversarial resilience. After the hybrid method and random filtering, binary filtering has the highest accuracy of 57%. The hybrid method improves the accuracy by 55% over the next best traditional defense method, which is random filtering.

**TABLE 7: Comparison of defense methods for C&W attack.**

| Defense Method | Precision | Recall | F1-Score | Accuracy |
|---|---|---|---|---|
| Inception-V3 | 0.24 | 0.13 | 0.15 | 0.13 |
| JPEG Filter | 0.89 | 0.87 | 0.87 | 0.87 |
| Feature Squeeze | 0.66 | 0.60 | 0.61 | 0.60 |
| Binary Filter | 0.79 | 0.67 | 0.70 | 0.67 |
| Random Filter | 0.92 | 0.89 | 0.90 | 0.89 |
| Hybrid | 0.92 | 0.89 | 0.90 | 0.89 |

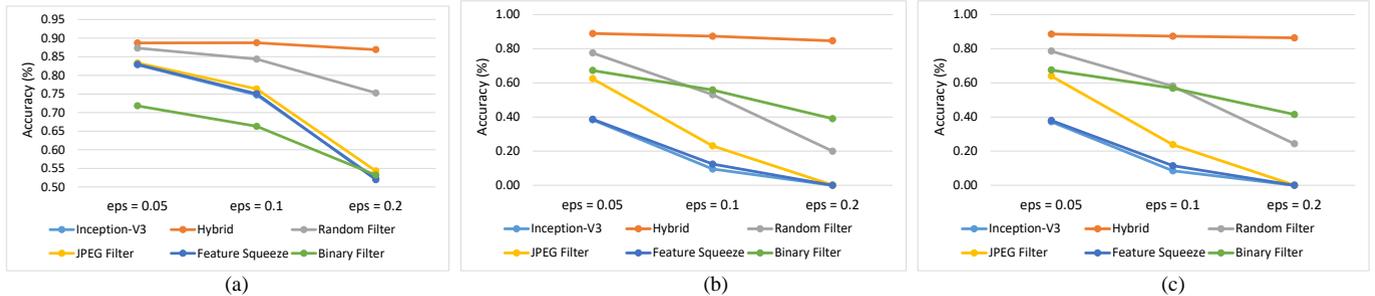

**FIGURE 7.** Performance variation of defense methods with ε for (a) FGSM Attack, (b) MIM Attack, (c) PGD Attack.

### 5) C&W ATTACK

For the C&W attack, we have used the $L_2$ norm and 80 maximum iterations. The results are presented in Table 7. The C&W attack is very effective in reducing the accuracy of the Inception-V3 model to 13%. The hybrid defense method performs best with the highest precision, recall, F1-score, and accuracy of 92%, 89%, 90%, and 89%, respectively. Random filtering achieves a high accuracy of 89%. Therefore, we can conclude that random filtering has a major impact in terms of adversarial resilience against C&W attacks. In the hybrid method, the first line of defense is local feature mapping, which is primarily responsible for identifying the traffic signs with text. For non-text signs, random filtering is able to nullify the effect of adversarial perturbations, so the Resnet-152 model has no impact on the identification of the correct traffic sign label. With random filtering, the Inception-v3 can achieve the same performance as the hybrid method. Therefore, for the C&W attack, we can conclude that random filtering and local feature mapping are the two dominant techniques in terms of adversarial resilience.

### B. PERFORMANCE EVALUATION WITH VARIATION OF PERTURBATION MAGNITUDE

We vary the value of ε for FGSM, MIM, and PGD attacks for their performance evaluation. Earlier studies used perturbation values between 0.05 to 0.2 [41], [42]. The perturbation values used in this study are 0.05, 0.1, and 0.2. For all three attacks, the hybrid method shows the most resilience compared to all other methods, as shown in Figure 7. Our analysis revealed that, for MIM and PGD attack models, an ε value of 0.2 reduces the traditional methods' accuracies from 99% close to 0%. The ε represents the level of distortion created on the input images by the perturbation model of the attack, so an ε value of 0.2 for MIM and PGD attacks creates enough distortions in the images such that the Inception-V3 model has a near 100% misclassification rate. However, if we use the hybrid defense method with the Inception-V3 model, for an ε value of 0.2 for MIM and PGD attack is unable to decrease the classification accuracy significantly, and the Inception-V3 model is able to classify traffic signs with an accuracy comparable to lower ε values of 0.1 and 0.05. Therefore, it can be concluded that the hybrid defense method is resilient against attacks with high ε values (i.e., with highly distorted input images).

### C. CLASS-WISE PERFORMANCE ANALYSIS OF HYBRID DEFENSE METHOD FOR PGD ATTACK

From Table 8, we observe that the hybrid defense method performs well across all 18 labels. The precisions value for "Do Not Enter" sign and "Pedestrian Crossing" sign is relatively lower, which means that the model sometimes misclassifies other traffic signs as these two signs. The "Lane ends" sign and "Roundabout" sign have lower recall values. These two signs have more variation in images compared to other traffic signs, which is the reason for the misclassifications related to these particular signs. Moreover, most of the Lane ends and roundabout signs do not contain any text, so the local feature mapping cannot contribute to these scenarios. Overall, the traffic signs with text have higher recall compared to non-text signs.

**TABLE 8:** Class-wise performance of hybrid defense method for PGD attack scenario.

| Traffic Sign Type | Precision | Recall | F1-Score | Number of Samples |
|---|---|---|---|---|
| Curve left | 0.91 | 1 | 0.95 | 39 |
| Curve right | 0.98 | 0.85 | 0.91 | 114 |
| Do not enter | 0.68 | 0.93 | 0.79 | 68 |
| Lane ends | 1 | 0.73 | 0.84 | 62 |
| Merge | 0.72 | 0.85 | 0.78 | 33 |
| Pedestrian crossing | 0.49 | 0.84 | 0.62 | 88 |
| Roundabout | 1 | 0.64 | 0.78 | 61 |
| School zone | 0.92 | 0.95 | 0.93 | 60 |
| Signal ahead | 0.72 | 1 | 0.84 | 39 |
| Speed limit 15 | 0.93 | 0.94 | 0.93 | 68 |
| Speed limit 30 | 1 | 0.82 | 0.9 | 110 |
| Speed limit 35 | 0.98 | 0.99 | 0.98 | 92 |
| Speed limit 45 | 1 | 0.96 | 0.98 | 107 |
| Speed limit 55 | 0.92 | 0.9 | 0.91 | 171 |
| Speed limit 65 | 0.99 | 0.73 | 0.84 | 134 |
| Stop | 0.91 | 0.97 | 0.94 | 79 |
| Stop ahead | 0.86 | 0.71 | 0.78 | 45 |
| Yield | 0.94 | 0.95 | 0.94 | 81 |

We plot the confusion matrix of the hybrid defense method under PGD attack with ε=0.1. In Figure 8, we can see the confusion matrix of the classification results. The speed limit signs are sometimes misclassified as pedestrian crossing signs. Moreover, within speed limit signs, the model sometimes classifies the 65 mph sign as 55 mph. This is due to the similarity between the digits 5 and 6.

Among the non-speed limit signs, the misclassification generally occurs in the "Do Not Enter" sign. Because of these issues, the precision of do not enter sign and pedestrian crossing sign is low. Overall, the our hybnrid defense method performs well across all types of traffic signs as observed in Figure 8.

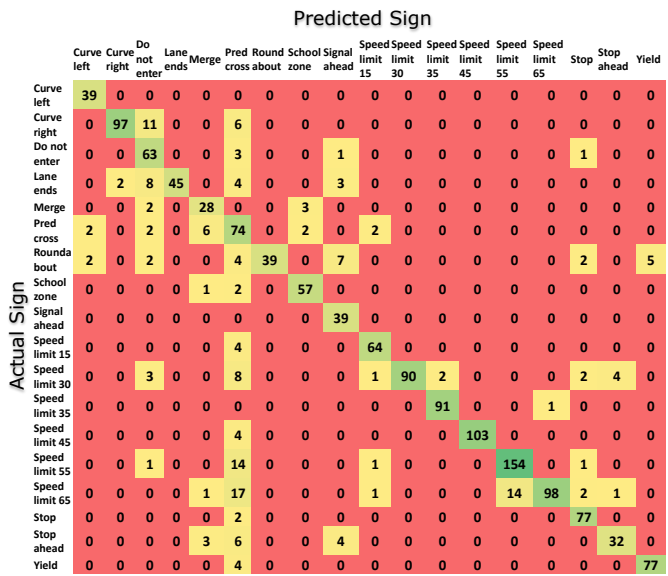

**FIGURE 8.** Confusion matrix of hybrid defense method for PGD attack scenario.

## VIII. CONCLUSIONS AND FUTURE WORK

In this study, we have developed a hybrid defense method against adversarial attacks on traffic sign classifiers used for the autonomous vehicle perception module. The DNN-based hybrid defense method developed is a combination of three different approaches; random filtering, ensembling and local feature mapping. Our analysis revealed that our hybrid defense method is resilient to different types of white-box adversarial attacks, as proven by the average classification accuracy of 88% achieved under different types of attacks, such as FGSM, MIM, PGD and C&W. When compared with traditional defense methods, such as JPEG filtering, feature squeezing, and binary filtering, our hybrid method shows better performance compared to these traditional defense methods for sign classification accuracy under adversarial attacks. The hybrid defense method against adversarial attacks on traffic sign classifiers is developed and evaluated in this study using computation units that have less computation capabilities than the computational units used in AVs as discussed earlier in the paper. Thus, we can use the hybrid defense method for real-world AV operations to classify adversarial attacks by embedding the hybrid defense method in an AV's in-vehicle computation unit.

Several areas for improvement of our hybrid method have been identified from the class-wise performance analysis of the hybrid method. For example, misclassifications between specific classes, such as 55 and 65 mph speed limit signs can be addressed in future research. Moreover, the hybrid method has only been tested for a specific application in this study, which is traffic sign recognition. It should also be tested for other benchmark datasets, such as the MNIST and CIFAR datasets which contain images about hand-written digits, vehicle, animal, etc.. Finally, our hybrid defense method should also be tested for other types of attacks beyond the attacks considered in this study. This study, however, is a good starting point for exploring hybrid approaches for adversarial resilience in traffic sign classifiers in autonomous vehicles.

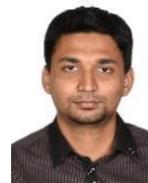

**Zadid Khan** received the B.Sc. degree in electrical and electronic engineering from Bangladesh University of Engineering and Technology (BUET), Dhaka, Bangladesh, in 2014. Then he served as a petroleum engineer in asset development department of Chevron BPC (Bangladesh Profit Center) from 2014 to 2016. After that, he received his M.Sc. degree in civil engineering (transportation major) from Clemson University, Clemson, USA, in 2018 and his PhD in the same department and University in 2021. Moreover, he joined the cyber physical systems (CPS) lab as a graduate research assistant from fall 2016, under the supervision of Dr. Mashrur Chowdhury, Professor, dept. of Civil Engineering. His primary research focus is connected and autonomous vehicles (CAVs). Within the CAV domain, his research interests are machine/deep learning, computer networking (V2X, SDN, HetNet), data analytics (data fusion, big data), cloud computing and cybersecurity.

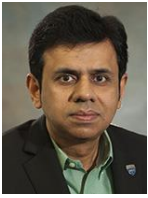
**Mashrur "Ronnie" Chowdhury** (SM'14) is the Eugene Douglas Mays Chair of Transportation at Clemson University. He is the director of the USDOT Center for Connected Multimodal Mobility (C2M2) ((http://cecas.clenson.edu/c2m2). He is the co-director of the Complex Systems, Analytics and Visualization Institute (CSAVI) (http://clemson-csavi.org) at Clemson University. He is the director of the Transportation Cyber-Physical-Social Systems Laboratory at Clemson University. He previously served as an elected member of the IEEE ITS Society Board of Governors and is currently a senior member of the IEEE. He is a Fellow of the American Society of Civil Engineers (ASCE) and an alumnus of the National Academy of Engineering (NAE) Frontiers of Engineering program. He is the founding advisor of the "IEEE Intelligent Transportation Systems Society (ITSS) Student Chapter" at Clemson University. He is a registered professional engineer in Ohio.

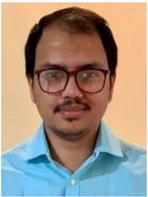
**Sakib Mahmud Khan** (Member'20) is the Assistant Research Professor at Glenn Department of Civil Engineering, Clemson University, and Assistant Director of the Center for Connected Multimodal Mobility ($C^2M^2$). Before joining the center, he was a post-doctoral research scholar working at California Partners for Advanced Transportation Technology (PATH), University of California Berkeley. He received his Ph.D. and M.Sc. in Civil Engineering from Clemson University in 2019 and 2015, respectively.